\documentclass[letterpaper, 10 pt, journal, twoside, print]{ieeecolor}
\usepackage{lcsys}
\usepackage{cite}
\usepackage{mathtools,amsthm,amsmath,mathalfa,bbm,amssymb}
\usepackage{color}
\usepackage{comment}
\usepackage{textcomp}
\makeatletter
\def\endfigure{\end@float}
\makeatother
\usepackage{subfig}

\renewenvironment{proof}{
\noindent{\sc Proof.}\hspace{0.10cm}\,\,}{$\hfill\Box$\vspace{.1cm}}

\newcommand{\dd}{\textrm{d}}

\usepackage{algorithmic}
\usepackage{graphicx}
\usepackage{textcomp}
\usepackage{subfig}
\usepackage{xspace}

\newcommand{\rline}{{\mathbb R}}

\newcommand{\nline}{{\mathbb N}}

\newcommand{\rfb}[1]{\mbox{\rm
		(\eref{#1})}\ifx\undefined\stillediting\else:\fbox{$#1$}\fi}

\newcommand{\bluff}{{\hbox{\raise 15pt \hbox{\hskip 0.5pt}}}}

\newfont{\roma}{cmr10 scaled 1200}

\newtheorem{thm}{Theorem}[section]

\newtheorem{prop}[thm]{Proposition}

\def\BibTeX{{\rm B\kern-.05em{\sc i\kern-.025em b}\kern-.08em
    T\kern-.1667em\lower.7ex\hbox{E}\kern-.125emX}}
\begin{document}
\title{Newton and Secant Methods for Iterative Remnant Control of Preisach Hysteresis Operators}
\author{J.R. Keulen, \IEEEmembership{Graduate Student Member, IEEE} , B.Jayawardhana, \IEEEmembership{Senior Member, IEEE}
 \thanks{This research project is supported by a TKI (Topconsortia voor Kennis en Innovatie) grant within the Top Sector High-Tech Systems and Materials (HTSM).}
 \thanks{J.R. Keulen and B. Jayawardhana are with the Engineering and Technology Institute Groningen, Faculty of Science and Engineering, University of Groningen, 9747AG Groningen, The Netherlands (e-mails: j.r.keulen@rug.nl, b.jayawardhana@rug.nl)}}

\maketitle
\thispagestyle{empty}
\begin{abstract}
We study the properties of remnant function, which is a function of output remnant versus amplitude of the input signal, of Preisach hysteresis operators. The remnant behavior (or the leftover memory when the input reaches zero) 
enables an energy-optimal application of piezoactuator systems where the applied electrical field can be removed when the desired strain/displacement has been attained.  
We show that when the underlying weight of Preisach operators is positive, the resulting remnant curve is monotonically increasing and accordingly a Newton and secant update laws for the iterative remnant control are proposed that allows faster convergence to the desired remnant value than the existing iterative remnant control algorithm in literature as validated by numerical simulation. 
\end{abstract}

\begin{IEEEkeywords}
Hysteresis, Preisach hysteresis operator, Remnant control, Mechatronics, Newton's method
\end{IEEEkeywords}

\section{Introduction}
\label{sec:introduction}
\IEEEPARstart{H}{Ysteresis} is a phenomenon where the response of a system depends not only on its current input but also on the past and present memory of its state. 
Hysteresis behaviors are commonly encountered in materials with memory, for example, in ferromagnetic, shape memory alloy, and piezo-electric systems. It is important to study hysteretic behavior to control such systems. 

There are multiple mathematical models discussed in the literature to describe hysteretic behavior.  Depending on whether the hysteresis behavior is affected by the rate of the input, one can have rate-independent hysteresis models and the rate-dependent ones.
Rate-dependent hysteretic behavior can, for example, be modeled by non-smooth integrodifferential equations, such as the Duhem models \cite{ikhouane2018survey}, although they have also been used to represent rate-independent hysteretic behavior. Other 
models are based on infinite-dimensional 
operators, which include the well-studied Preisach operators. The Preisach operators are constructed from an infinite number of hysterons, which are typically given by 
relay operators  \cite{mayergoyz2003mathematical}. There are a number of variations of the Preisach operators, such as 
the Krasnosel’skii–Pokrovskii (KP) operators 
\cite{krasnosel2012systems},  
the Prandtl-Ishlinskii 
\cite{al2018internal} and many others. We refer the interested readers to the exposition on hysteresis models in 
\cite{hassani2014survey, visintin2013differential, brokate1996hysteresis}.

Various control methods for hysteretic systems are discussed in the literature. The general objective is to employ hysteresis models, such as the standard  Preisach operators, to accurately represent the system's behavior. Subsequently, the inverse of this hysteresis model is integrated into feedforward control methodologies to compensate for the unwanted hysteretic behavior \cite{iyer2005approximate} or to use an internal property of hysteresis, such as, energy dissipation \cite{jayawardhana2012} or sector-bound conditions to conclude systems' stability \cite{jayawardhana2008, jayawardhana2009}. In 
\cite{al2018internal}, the authors  utilized the rate-dependent Prandl-Ishlinskii hysteresis model to design 
a 
nonlinear controller containing 
an inverse multiplicative structure of the hysteresis model. By studying the time-derivative of input and output of hysteresis operators as in \cite{jayawardhana2008,logemann2003,tarbouriech2014}, feedback stability of hysteretic systems can be concluded via absolute stability analysis. Feedback control analysis and design that relies on the inherent energy dissipativity property of general hysteresis operators (given by either the Preisach or Duhem models) is presented in 
\cite{gorbet2001passivity,ouyang2014}. An inversion-free feedforward hysteresis control approach is found in \cite{ruderman2023inversion}. In all these works, the output regulation of a constant reference point requires a constant control input to the (hysteretic) actuator. Consequently, a constant consumption of power is needed and it can lead to engineering design difficulty when a large number of such actuators are used for particular high-tech applications, such as, the high-density deformable mirror proposed in  \cite{huisman2021high,schmerbauch2020}.  


In contrast to having an active control for achieving output regulation of hysteretic systems as above, we study in this paper the problem of output remnant control where the control input can be set to zero when the desired output has been reached; thanks to the hysteresis' remnant property. Generally speaking, the output remnant is the leftover memory of the hysteresis when the input is set to zero. Depending on the input history or memory of the hysteresis, the output remnant can take any value from an admissible remnant interval. In this regard, the output remnant control corresponds to designing an admissible input signal that can bring the output remnant from an initial remnant value to a desired remnant state. 
As described before, the control of this remnant value is relevant for applications that require minimal use of input control due to various factors, such as, complex power electronics design for controlling high-density hysteretic actuators, power constraint, 
or the consequent energy loss/heat dissipation associated with the use of a constant non-zero input to sustain the desired output. An example of such application in the high-precision opto-mechatronics systems 
that exploit such output remnant behaviors is the development of hysteretic deformable mirror for space application \cite{huisman2021high}. The deformable mirror uses a Nb-doped piezo material (PZT), which allows a wide range of remnant deformation \cite{jayawardhana2018modeling}. In order to address the remnant control problem in these applications, a recursive remnant control algorithm has been proposed in \cite{vasquez2020recursive} that is based on a standard iterative learning control law where the input amplitude is updated proportional to the remnant error. The learning rate / gain affects the convergence rate of the remnant control and it cannot be set arbitrary fast without inducing instability. 

In this paper, we study the remnant property of Preisach hysteresis operators that will allow us to design an output remnant control law with better transient performance. Firstly, we investigate the property of remnant curve of such operators as a function of input amplitude for a given initial memory state. Secondly, based on the continuity and monotonicity property of the remnant curve, we propose a new iterative remnant control algorithm based on Newton and Secant methods. 

The rest of this paper is structured as follows. In Section \ref{sec:II} we introduce the Preisach operators and the remnant control problem formulation. 
In Section \ref{sec:III}, we  study some mathematical properties of the {remnant curve} of the Preisach operators. Subsequently, a new iterative remnant control algorithm is introduced in Section \ref{sec:ffremnant} and its numerical validation is presented in 
Section \ref{sec:simulation}.  

\section{Preisach hysteresis operators and remnant control problem}\label{sec:II}

We denote $C(U, Y)$, $AC(U, Y)$, and $C_{pw}(U, Y)$ as the sets of continuous, absolutely continuous, and piece-wise continuous functions \(f: U \to Y\), respectively.  The space of $p$-integrable measurable functions $f:U\to Y$ is denoted by $\mathcal L^p(U,Y)$. The Sobolev space $W^{k,p}(U,Y)$ is a subset of measurable functions $f\in L^p(U,Y)$ where its weak-derivatives up to order $k$ belong also to $L^p(U,Y)$.
\subsection{Preisach hysteresis operator}
Let us present a formal definition of standard Preisach operators, as presented in \cite{mayergoyz2003mathematical}. For this purpose, we firstly define the Preisach plane $P$ by  $P:=\big\{(\alpha,\beta)\in \mathbb{R}^{2}|\alpha\geq\beta\big\}$. In this plane, an interface line can be defined that separates the relays of Preisach operator (which will be defined shortly) which have positive values and the ones with negative values. Specifically, an interface line $L$ is defined by monotonically increasing sequences $\alpha_i\geq 0$ and $\beta_i\geq 0$, $i\in\nline$ with $\alpha_1=\beta_1=0$, and is given by 
\begin{equation}\label{eq:interface}
L:=\left\{
\begin{array}{ll}\bigcup_{i\in\nline}\left( [\alpha_{i},\alpha_{i+1}]\times \{-\beta_i\}\right) \cup \left(\{\alpha_{i+1}\}\times [-\beta_{i+1},-\beta_{i}]\right) \ \text{or} \\
\bigcup_{i\in\nline}\left( \{\alpha_{i}\}\times [-\beta_{i+1},-\beta_{i}] \right) \cup \left( [\alpha_{i},\alpha_{i+1}]\times \{-\beta_{i+1}\}\right).
\end{array}\right.
\end{equation}
Roughly speaking, the interface is a staircase line that starts from $(0,0)$ with horizontal or vertical sub-lines, and is defined only in the second quadrant of the Preisach plane $P$, i.e. in $\rline_-\times \rline_+=:P_2$.  
Let us denote by $\mathcal{I}$ the set of all interface lines $L\in P_2$. 
For a given initial interface $L_{0}$,  
the Preisach operator $\mathcal{P}:AC(\mathbb{R}_{+},\mathbb{R})\times\mathcal{I}\rightarrow AC(\mathbb{R}_{+},\mathbb{R})$ can be formally defined by
\begin{equation}\label{eq:Preisach}
    \big(\mathcal{P}(u, L_{0})\big)(t):=\iint\limits_{(\alpha, \beta)\in P}w(\alpha,\beta)\big(\mathcal{R}_{\alpha, \beta}(u,L_{0})\big)(t)\dd\alpha \dd \beta
\end{equation}
where $w:\mathbb{R}\rightarrow\mathbb{R}_{+}$ is the weight function, 
and 
$\mathcal{R}_{\alpha, \beta}(u,L_0)$ is the relay operator defined by
\begin{equation*}\label{eq:relay}
        \Big(\mathcal{R}_{\alpha, \beta}(u)\Big)(t):=
        \begin{cases}
        1 & \text{if} \ u(t)>\alpha,\\
        -1 & \text{if} \ u(t)<\beta,\\
        \Big(\mathcal{R}_{\alpha, \beta}(u)\Big)(t_{-}) & \text{if} \ \beta\leq u(t)\leq\alpha,  t>0,\\
        r_{\alpha,\beta}(L_0) & \text{if} \ \beta\leq u(t)\leq\alpha,  t=0,
        \end{cases}
\end{equation*}
where we omit the argument of $L_0$ in the definition for conciseness and $r_{\alpha,\beta}(L_0)$ is the initial state of the relay $\mathcal R_{\alpha,\beta}$ which is equal to $-1$ if $(\alpha,\beta)$ is located above $L_0$ and $1$ otherwise.    


For $h\geq0$ and $\mathbb{R}_{h}:=[-h,\infty)$, a function $f:\mathbb{R}_{h}\rightarrow\mathbb{R}_{h}$ is a \textit{time transformation} if $f$ is continuous and non-decreasing with $f(-h)=-h$ and $\lim\limits_{t\rightarrow\infty}f(t)=\infty$; in other words $f$ is a time transformation if it is continuous, non-decreasing and subjective. Following the work of \cite{logemann2008class}, the Preisach operator $\mathcal{P}$ is \textit{rate independent}, i.e. for every time transformation $f$ it holds that
\begin{equation}\label{eq:rateindependent}
    (\mathcal{P}(u \circ f))(t)=(\mathcal{P}(u))(f(t)), \ \forall \ u\in C(\mathbb{R}_{h}), \ \forall t\in\mathbb{R}_{h}.
\end{equation}

\subsection{Remnant control problem for Preisach operators}

In general, for any given desired remnant position $y_d$, the remnant control problem pertains to the design of input signal $u$, where $u(t) = 0$ for all $t>T$ for some $T>0$ and $u(t)\neq 0$ on $(0,T)$, such that the corresponding output of the hysteresis operator will be equal to $y_d$ after the input is set to zero (i.e. after time $T$). For the Preisach operator with a given initial interface $L_{0}\in\mathcal{I}$,  
the application of such input signal $u$ 
should lead to $\mathcal{P}(u,L_{0})(t)=y_d$ for all $t\geq T$. Intrinsically, the input signal $u$ alters the interface function such that some relays in the Preisach domain $P$ have switched from the initial state $r_{\alpha,\beta}$ so that the remnant value is equal to $y_d$. 

Let $L_{T}\in\mathcal{I}$ be the final interface that describes the state of the relays in the Preisach operator at $t=T$ and correspondingly, we define $P_{T}\subset P$ be the domain of relays $\mathcal R_{\alpha,\beta}$ that have switched their value from their initial condition $r_{\alpha,\beta}$ that depends on $L_0$. 
Using $P_T$, the challenge of remnant control is on the design of 
a feedforward control input $u$ such that its values are zero at a given terminal time $T$ and the remnant output that is due to the switched relays in $P_T$ is equal to the required incremental output needed to bring the initial output $y_0$ to the desired one $y_d$. 

In \cite{vasquez2020recursive} an iterative algorithm is proposed based on an input of the form
\begin{equation}\label{eq:iterative_remnant_input}
    u(t):=\sum\limits_{k=0}^{\infty}A_{k}v_k(t),
\end{equation} 
where $v_{k}$ is a triangular pulse signal with a unit amplitude whose pulse starts at $t=kT$ and vanishes at $t=(k+1)T$ with $T$ be the periodic remnant update time and $A_k$ be the amplitude for the $k$-th pulse. 
The input $u$ can be seen as a sequence of triangular pulses whose amplitudes are modulated by  $A_{k}$. By the application of such input signal, the remnant output after the application of the $k$-th triangular wave $A_kv_k(t)$ is given by 
\begin{equation}
    \begin{split}
        y_k:&=\Big(\mathcal{P}(u,L_{0})\Big)((k+1)T)
=\Big(\mathcal{P}(A_{k}v_{0},L_{k})\Big)(T).
    \end{split}
\end{equation}
As studied in \cite{vasquez2020recursive}, the iterative remnant control design problem corresponds to the design of an update rule for the amplitude $A_{k+1}$ based on the current amplitude $A_k$ and remnant value $y_k$ 
such that $y_k$ converges to $y_{d}$ as $k\rightarrow\infty$. Particularly, the update rule that is studied in \cite{vasquez2020recursive} is given by 
\begin{equation}\label{eq:iterative_remnant}
    A_{k+1}=A_{k}-\lambda e_{k},
\end{equation}
where $e_{k}=y_k-y_{d}$ is the remnant error and $\lambda$ is the adaptation gain. A bound on the adaptation gain $\lambda$ is further studied in \cite{vasquez2020recursive} that guarantees the convergence of $A_k$ in \eqref{eq:iterative_remnant} to the desired amplitude for attaining the desired remnant position. This convergence property relies on 
a monotonicity property of the remnant position as a function of the input amplitude $A$.  This method relies on a constant gain value, which can limit the convergence speed to the desired remnant position. With the introduction of the remnant curve, where we investigate the mapping between the amplitude $A$ and the remnant value, we can introduce a new remnant control algorithm that exploits an adaptive gain in the form of the Newton method. 



\section{Remnant curve and its properties}\label{sec:III}

Let us firstly define the remnant curve $\rho_{L_{0}}:\mathbb{R} \rightarrow \mathbb{R}$, that maps the input amplitude (defined shortly below) to the 
remnant position. 
Note that the remnant curve, in this case, is dependent on the initial interface line $L_0$. For a different initial interface line, the remnant curve will be different. 
Let $v(A,t):\mathbb{R}\times \mathbb{R}\rightarrow \mathbb{R}$ be a piecewise-continuous input parametrized by the amplitude $A$ such that there exist a $T_{2}>T_{1}>0$ so that the following conditions hold:
\begin{description}
\item[R1.] $v(A,0)=0$ and  $v(A,t)=0$ for all $t\geq T_{2}$; 
\item[R2.] $v(A,t)$ is monotone on each time interval $t\in(0,T_{1})$ and $t\in (T_{1},T_{2})$;  
\item[R3.] $\frac{d}{dt}v(A,T_{1}) \frac{d}{dt}v(A,T_{2})<0$ for all $T_1 \in (0,T_{1})$ and $T_{2}\in(T_{1},T_{2})$; 
\item[R4.] $\text{sign}(\frac{d}{dt}v(A,T_{1}))=\text{sign}(A)$.
\end{description}
Using the above input signal $v_\rho(A,t)$, the remnant curve $\rho_{L_0}$ can then be defined by 
\begin{equation}\label{eq:rho_L_01}
\rho_{L_0}(A) = \lim_{t\to\infty}(\mathcal P(v(A,\cdot),L_0))(t).
\end{equation}
Note that we use the asymptotic value in the above definition so that it is independent of a particular value of $T_2$, which may not be unique. One can also replace the asymptotic value in the above definition if one fixes the value of $T_2$ when we design the {\it feedforward remnant input} as will be used later in Section IV. In this case, we can modify \eqref{eq:rho_L_01} into
\begin{equation}\label{eq:rho_L_02}
\rho_{L_0}(A) = (\mathcal P(v(A,\cdot),L_0))(T),
\end{equation}
where $T$ is the time at which $v(A,T)=0$. The initial remnant value is trivially given by $\rho_{L_0}(0)$. Due to the rate-independent property of the Preisach operator, one can immediately check that $\rho_{L_0}$ is invariant to the particular form of $v_\rho$ as shown in the following proposition.

\begin{prop}\label{prop:remnant_curve_invariant}
Consider a Preisach operator $\mathcal P$ as in \eqref{eq:Preisach} with a given weight function $w$ and initial interface $L_0$. For any initial interface $L_0$, the corresponding remnant curve $\rho_{L}$ is 
invariant to the particular form of $v(A,t)$ satisfying R1 -- R4.
\end{prop}

\begin{proof}
We will prove this proposition by showing that for any pair of $v_{1}(A,t)$ and $v_{2}(A,t)$, we have 
\[
\lim_{t\to\infty}(\mathcal P(v_{1}(A,\cdot),L_0))(t) = \lim_{t\to\infty}(\mathcal P(v_{2}(A,\cdot),L_0))(t)
\]
for all $A\in \rline$. 

As shown before, the Preisach operator has rate-independent property, namely, for all time transformation $f$, it holds that 
\begin{equation}\label{eq:P_v1_v2}
\Big(\mathcal P(v_{1}(A,\cdot)\circ f)\Big)(t) = \Big(\mathcal P(v_{1}(A,\cdot))  \Big)(f(t)).
\end{equation}
By the monotonicity of $v_1$ on the time intervals $(0,T_{1,v_1})$ and $(T_{1,v_1},T_{2,v_1})$ and the monotonicity of $v_2$ on the time intervals $(0,T_{1,v_2})$ and $(T_{1,v_2},T_{2,v_2})$, we can define a time transformation $f$ for every $A$ such that $v_1(A,f(t)) = v_2(A,t)$ for all $t\geq 0$. In this case, \eqref{eq:P_v1_v2} becomes
\[
\Big(\mathcal P(v_{2}(A,\cdot))\Big)(t) = \Big(\mathcal P(v_{1}(A,\cdot))  \Big)(f(t)).
\]
By taking $t\to\infty$, we establish our claim. 
\end{proof}

 Proposition \ref{prop:remnant_curve_invariant} leads to the conclusion that the amplitude of the input signal serves as the variable parameter in the remnant control problem. The subsequent proposition follows from the wiping-out property of the Preisach hysteresis model and discusses how the interface evolves after the application of such input signal.
\begin{prop}\label{prop:wipingout}
Consider a Preisach operator $\mathcal P$ as in \eqref{eq:Preisach} with a given weight function $w$ and initial interface $L_0$ whose staircase line is defined by the sequences $\{\alpha_{i,0}\}_{i\in\nline}$ and $\{\beta_{i,0}\}_{i\in\nline}$ as in \eqref{eq:interface}. Then the new interface line $L_1$ (with the sequences $\{\alpha_{i,1}\}_{i\in\nline}$ and $\{\beta_{i,1}\}_{i\in\nline}$) after the application of input signal $u(t)=v(A,t)$ to $\mathcal P$ satisfies $\alpha_{1,1}=\beta_{1,1}=0$ and 
\begin{itemize}
\item if $A>\alpha_{2,0}$ then 
\begin{align*}
\alpha_{2,1} & = A, \qquad \alpha_{i,1}  = \alpha_{i+k_1-3,0} \qquad \forall i\geq 3 \\
\beta_{i,1} & = \beta_{i+k_2,0} \qquad \qquad \qquad \qquad \, \, \forall i\geq 2
\end{align*}
\item otherwise (i.e. if $A < -\beta_{2,0}$)
\begin{align*}
\alpha_{i,1}  & = \alpha_{i+k_3,0} \qquad \qquad \qquad \qquad \, \, \forall i\geq 2 \\
\beta_{2,1} & = -A, \qquad \beta_{i,1} = \beta_{i+k_4-3,0} \qquad \forall i\geq 3,
\end{align*}
\end{itemize}
where $k_2, k_3$ are the indices of $\alpha_{i,0}$ and $\beta_{i,0}$ such that $(A,-\beta_{i+k_2,0})\in L_0$ and $(\alpha_{i+k_3,0},-A)\in L_0$, respectively, and $k_1, k_4$ are the smallest indices of $\alpha_{i,0}$ and $\beta_{i,0}$ such that 
$\alpha_{k_1,0} > |A|$ and $\beta_{k_4,0} > |A|$.   
\end{prop}

\begin{proof}
We start the proof by analyzing the changes to the interface line when $A>\alpha_{2,0}$. The other case can be proven in a similar fashion. 

Firstly, when the input signal has reached its maximum value of $A$, all relays $\mathcal R_{\alpha,\beta}$ with $\alpha<A$ will switch to $1$. This implies that the horizontal and vertical lines of the interface that are associated to the sequence $\alpha_{i,0}<A$, $i\neq 0$ will be wiped out in the new interface and it creates a new horizontal line with $\alpha_{2,1} = A$. The rest of the new sequence $\alpha_{i,1}$ with $i>1$ will follow the sequence from the original one that is not wiped out, i.e., those associated to $\alpha_{i+k_1-1,0}$. Subsequently, when the input signal returns back to $0$, all relays $\mathcal R_{\alpha,\beta}$ with $\beta>0$ will be at $-1$ state and the vertical lines of the interface will be redrawn to the vertical line of $L_0$ that contains the element $(A,-\beta_{i,0})$ for some $i>1$ and all $\beta_{j,0}$ s.t. $\beta_{j,0}<\beta_{i,0}$ will be wiped out. 
\end{proof}

In Proposition \ref{prop:wipingout}, the removal of elements in the sequence of $\alpha$ and $\beta$ from the original one when the input $v(A,t)$ is applied to the Preisach operator is known in the literature as the {\it wiping-out} property. The characterization of this wiping-out property will be useful to get the relationship between the remnant value and the weights of the Preisach operator associated to the wiped-out domain. 


\begin{figure}[h]
    \centering
    \includegraphics[width=0.4\textwidth]{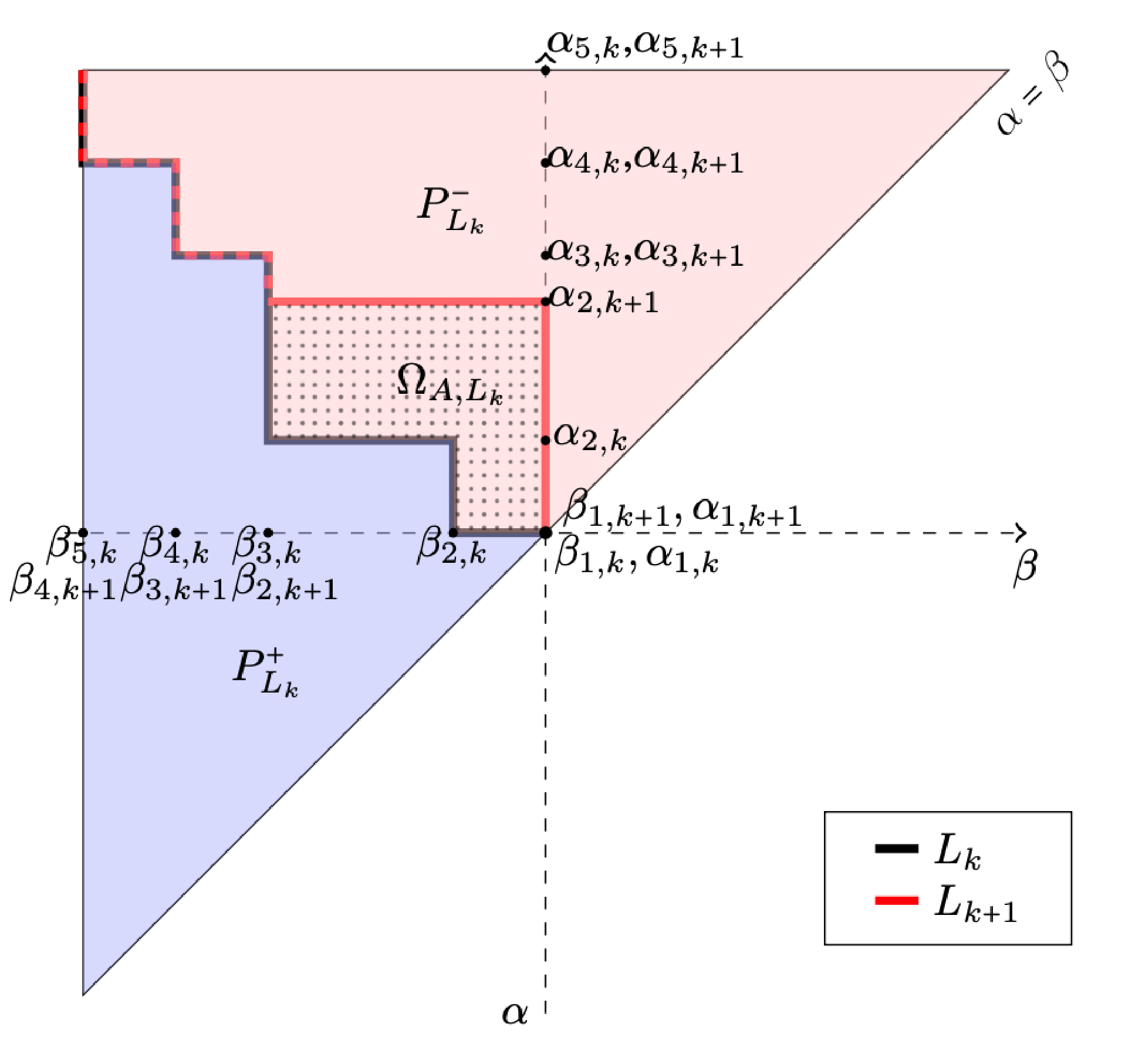}
    \caption{Preisach domain with a particular staircase interface $L_0$, consisting of the sequence $\{\alpha_{1,k},\  \alpha_{2,k},\ \alpha_{3,k},\ \alpha_{4,k},\ \alpha_{5,k}\}$ and $\{\beta_{1,k}$,  $\beta_{2,k}$, $\beta_{3,k}$, $\beta_{4,k}$, $\beta_{5,k}\}$, after application of input signal with $A_k=\alpha_{2,k+1}$ some horizontal and vertical lines of $L_k$ are wiped out, resulting in the new interface $L_{k+1}$ consisting of the sequence $\{\alpha_{1,k+1}, \ \alpha_{2,k+1},\ \alpha_{3,k+1},\ \alpha_{4,k+1},\ \alpha_{5,k+1}\}$ and $\{\beta_{1,k+1},\ \beta_{2,k+1},\ \beta_{3,k+1},\ \beta_{4,k+1}\}$. }
    \label{fig:PreisachDomain}
\end{figure}
Figure \ref{fig:PreisachDomain} shows an example of vertices being wiped out after an application of an input signal $v(A_k, t-kT)$, where amplitude $A_{k}=\alpha_{2,k+1}$. We can define a new region between $L_k$ and $L_{k+1}$, where the relay switched from -1 to +1. For the rest of this paper, we will denote this region by $\Omega_{A,L_k}$. 

\begin{prop}\label{prop:remnant_curve_explicit}
Consider a Preisach operator $\mathcal P$ as in \eqref{eq:Preisach} with a given weight function $w$, initial interface $L_k$ and the input signal $v(A,t)$ satisfying R1-R4 with $T_2=T$. 
Then the corresponding remnant curve $\rho_{L_k}$ is given by
\begin{equation}\label{eq:weightsremnantcurve}
\rho_{L_k}(A) = \iint\limits_{(\alpha, \beta)\in \Omega_{A,L_{k}}}w(\alpha,\beta)\dd \alpha \dd \beta + \rho_0,
\end{equation}
where region $\Omega_{A, L_{k}}$ is defined as the domain in $P$ that is enclosed by $L_k$ and $L_{k+1}(A)$ as given in Proposition \ref{prop:wipingout} and $\rho_0\in \rline$ is the initial remnant value due to 
$L_k$. 
\end{prop}

\begin{proof}
As shown in Proposition \ref{prop:remnant_curve_invariant} the remnant curve is invariant to the particular form of $v(A,t)$. 
Applying such input signal 
to $\mathcal P$ and using \eqref{eq:Preisach}, we can compute \eqref{eq:rho_L_02} as follows 
\begin{equation}\label{eq:integrals}
    \begin{split}
    \rho_{L_{k}}(A)=
    \iint\limits_{\beta \ \alpha} w(\alpha, \beta)\mathcal{R}_{\alpha,\beta}(v(A,\cdot))(T)\,\,\dd\alpha \, \dd \beta 
    \end{split}
\end{equation}
where we have used 
$\mathcal{R}_{\alpha, \beta}(v(A,\cdot))(0)=r_{\alpha,\beta}(L_k)$ describing the initial state of the relay $\mathcal R_{\alpha,\beta}$ which is equal to $-1$ if $(\alpha,\beta)$ is located above $L_k$ and $1$ otherwise. 
Let $P^{-}_{L_k}$ and $P^{+}_{L_k}$ denote the subdomains of the Preisach domain $P$ that are above or below the interface $L_k$. Let $L_{k+1}$ denote the new interface of the Preisach plane after the application of input $v(A,t)$. 
Then the first term on the RHS 
of \eqref{eq:integrals} can be expressed as
\begin{align*}
& \iint\limits_{\beta \ \alpha}w(\alpha, \beta)\mathcal{R}_{\alpha,\beta}(v(A,\cdot))(T)\,\dd\alpha \, \dd \beta  = -\iint\limits_{P^{-}_{L_k}}w(\alpha,\beta) \, \dd \alpha \, \dd \beta \\
& \qquad + \iint\limits_{P^{+}_{L_k}}w(\alpha,\beta) \, \dd \alpha \, \dd \beta 
+\iint\limits_{(\alpha, \beta)\in \Omega_{A,L_{k}}}w(\alpha,\beta) \, \dd \alpha \, \dd \beta, 
\end{align*}
where the last term corresponds to the fact that the relays in $\Omega_{A,L_k}$ have changed their state from $-1$ to $1$ due to the application of $v(A,t)$, and the sum of the first two terms is equal to the Preisach output when the interface $L_k$ is used, i.e. it is equal to initial remnant value $\rho_0$. 
\end{proof}

By utilizing the explicit definition of the remnant curve in \eqref{eq:weightsremnantcurve}, we can compute, under a specific assumption, the first derivative of $\rho_{L_k}(A)$ as a function of $A$. With this expression, we can study the sensitivity of 
the remnant curve with respect to the changes in the input signal amplitude. 

\begin{prop}\label{prop:derivative_rho} Let $L_{k}$ be an initial interface of $\mathcal P$ whose weight function $w\in C(P,\rline)\bigcap \mathcal L^1(P,\rline)$. 
Then $\rho_{L_{k}} \in W^{1,\infty}(P,\rline)$ (i.e. it is Lipschitz continuous). 
\end{prop}

\begin{proof}
It follows from Proposition \ref{prop:remnant_curve_explicit} that $\rho_{L_k}(A)$ is given by \eqref{eq:weightsremnantcurve}.  Following the analysis of hysteresis operator as in \cite{jayawardhana2012}, let us define a weak-derivative of $\rho_{L_k}$ using the upper-right Dini's derivative as follows
\begin{align*}
 \frac{\dd}{\dd A}\rho_{L_k}(A)  & := \limsup_{h\to 0^+} \frac{\rho_{L_k}(A+h)-\rho_{L_k}(A)}{h} \\ &
= \limsup_{h\to 0^+} \frac{\iint\limits_{(\alpha, \beta)\in \Delta_h}w(\alpha,\beta)\dd \alpha \dd \beta}{h},
\end{align*}
where the last equation is due to \eqref{eq:weightsremnantcurve} and $\Delta_h$ is the domain defined by $\Omega_{A+h,L_{k}}-\Omega_{A,L_{k}}$. 
Firstly, when $A$ is not at the boundary of the vertices of $L_k$, then it follows from the above equation and continuity of $w$ that for $A>0$, 
\begin{align*}
\frac{\dd}{\dd A}\rho_{L_k}(A) & = \limsup_{h\to 0^+} \frac{\int_{-\beta_{i,k}}^{0}\int_{A}^{A+h}w(\alpha,\beta)\dd \alpha \dd \beta}{h} \\
& = \int_{-\beta_{i,k}}^0 w(A,\beta)\dd \beta,
\end{align*}
with $i>1$ be s.t. $(A,-\beta_{i,k})\in L_k$, holds. Similarly, for $A<0$,
\begin{align*}
\frac{\dd}{\dd A}\rho_{L_k}(A) & = \limsup_{h\to 0^+} \frac{\int_{A}^{A+h}\int_{0}^{\alpha_{i,k}}w(\alpha,\beta)\dd \alpha \dd \beta}{h} \\
& = \int_{0}^{\alpha_{i,k}} w(\alpha,A)\dd \alpha
\end{align*}
with $i>1$ be s.t. $(\alpha_{i,k},-A)\in L_k$ holds. 
For this case, one can also check that taking the other limits of Dini's derivative (lower-right, upper-left and lower-left) 
also leads to the same quantity as above. 

On the other hand, when $A$ is equal to one of 
the interface vertices (i.e., $\alpha_{i,k}$ or $-\beta_{i,k}$ with $i\geq 2$), the computation of upper-right Dini's derivative leads to
\begin{itemize}
\item For $A>0$: $\frac{\dd}{\dd A}\rho_{L_k}(A)  =  \int_{-\beta_{i+1,k}}^0 w(A,\beta)\dd \beta$; 
\item For $A<0$: $\frac{\dd}{\dd A}\rho_{L_k}(A)  =  \int_{0}^{\alpha_{i+1,k}} w(\alpha,A)\dd \alpha$, 
\end{itemize}
with the same $i$ as before. The jump of the derivative at the interface vertices is due to the discontinuity of the staircase interface function $L$. 
Since $w$ is also $\mathcal L^1$, we can conclude that $\rho_{L_k}$ is Lipschitz continuous.  
\end{proof}



\begin{prop}\label{prop:monotonicity}
If $w$ is positive definite then for any given interface $L_k$ the remnant curve $\rho_{L_k}$ is monotone increasing. 
\end{prop}
 The proof follows directly from \eqref{eq:weightsremnantcurve} where a growing $|A|$ implies also that the region $\Omega_{A,L_k}$ enlarges. Hence $\rho_{L_k}$ is monotonically increasing. 
 
\section{Newton \& Secant Methods}\label{sec:ffremnant}
Let us introduce iterative remnant input signals given by \begin{multline}\label{eq:general_iterative_remnant_input}
u(t) = 
v(-A_{\text{max}},t) \\ + 
\sum_{k=1}^\infty{v(A_{k},t-(2k-1)T) + v(-A_{\text{max}},t-2kT)},
\end{multline}
where $v(A,t)$ is the input signal modulated by $A$ and satisfies R1-R4 with $T_2=T$, and the constant $A_{\text{max}}>0$ is a maximum remnant input amplitude. 
The first term and last term in \eqref{eq:general_iterative_remnant_input} are reset sub-signals 
such that the interface at each iteration step $k$ will be reset to $L_k=(\{0\}\times [-A_{\text{max}},0]) \cup ([0,A_{\text{max}}]\times\{-A_{\text{max}}\})\cup \ldots$. Using this iterative remnant input signal, our main design problem is to define an update law for $A_k$ such that $A_k$ converges to the desired amplitude where the remnant value is equal to the desired one. Due to the particular reset sub-signals in \eqref{eq:general_iterative_remnant_input}, it is implicitly assumed that the desired remnant output $y_d$ is larger than the remnant output after the initial reset signal but is still within the admissible range so that there exists $0<A_d< A_{\text{max}}$ such that $\rho_{L_1}(A_d)=y_d$. However, if $y_d$ is smaller than the remnant output after the initial reset signal, we can reverse the sign of reset sub-signals into $v(A_{\text{max}},t)$ and $v(A_{\text{max}},t-2kT)$ correspondingly, so that the target amplitude will satisfy $A_d<0$. In both cases, we need to design an iterative remnant control such that $A_k\to A_d$. In the following analysis, we will consider the first case where we use \eqref{eq:general_iterative_remnant_input} with $0<A_d<A_{\text{max}}$. 
For every iteration step $k$, define the remnant error $e_k$ (as in \eqref{eq:iterative_remnant}) by 
$e_k = \rho_{L_k}(A_k) - y_d$. 

\begin{thm}\label{thm:Newton_remnant}
Consider a Preisach operator $\mathcal P$ as in \eqref{eq:Preisach} with a positive-definite 
 $w\in C^1(P,\rline_+)$ and an initial interface $L_0$. 
Consider an iterative remnant input signal $u$ as in \eqref{eq:general_iterative_remnant_input} with $A_{\text{max}}>0$, and let $y_d>\rho_{L_1}(0)$ be the desired remnant value with $L_1$ be the interface after the first reset sub-signal. 
Let $A_k$ be updated by the following Newton's iterative remnant update law
\begin{equation}\label{eq:Newton_method}
A_{k+1} = A_k - \frac{e_k}{\rho_{L_k}'(A_k)},
\end{equation}
where $\rho_{L_k}'(A_k)$ is the derivative of $\rho_{L_k}$ as in Proposition \ref{prop:derivative_rho}. 
Then for an initial amplitude $A_1$ sufficiently close to $A_d$, $e_k$ converges quadratically to zero as $k\to\infty$ i.e. for some $\delta>0$,
\begin{equation}\label{eq:quadratic_conv}
|A_{k+1}-A_d| \leq \delta |A_k-A_d|^2.
\end{equation}

\end{thm}
\vspace{-0.5cm}\begin{proof}
Due to the introduction of reset sub-signals in \eqref{eq:general_iterative_remnant_input}, 
the remnant curve $\rho_{L_k}$ will be identical for every iterative step $k$ within the interval $\mathcal I:=[0,A_{\text{max}})$ according to Proposition \ref{prop:wipingout} and we will simply denote it by $\rho_L$. Following the proof of Proposition \ref{prop:derivative_rho}, the remnant curve $\rho_{L}$ is $C^1$ in $\mathcal I$ since its derivative is continuous and it is monotonically increasing following Proposition \ref{prop:monotonicity}. Particularly, $\frac{\dd}{\dd A}\rho_L(A)=\int_{-A_{\text{max}}}^0 w(A,\beta)\dd\beta>0$ and $\frac{\dd^2}{\dd A^2}\rho_L(A)=\int_{-A_{\text{max}}}^0 \frac{\dd}{\dd A}w(A,\beta)\dd\beta$. Moreover, since $w\in C^1$, 
we also have that $\rho_{L}\in C^2$. 

Let us denote $E_k=A_d-A_k$. 
 By Taylor series, we have
\[
0 = \rho_L(A_k) - y_d + \frac{\dd}{\dd A}\rho_L(A_k)(E_k)+\frac{\dd^2}{\dd A^2}\rho_L(\xi)(E_k)^2,  
\]
where $\xi\in [A_k,A_d]$. It follows from this equation that 
\begin{align*}
\frac{\rho_L(A_k) - y_d}{\rho_L'(A_k)} + A_d - A_k & = - \frac{\frac{\dd^2}{\dd A^2}\rho_L(\xi)}{\frac{\dd}{\dd A}\rho_L(A_k)}(A_d-A_k)^2.
\end{align*}
Now substracting both sides in \eqref{eq:Newton_method} by $A_d$ and substituting the above relation to the RHS term of \eqref{eq:Newton_method}, we arrive at \\ 
$|A_{k+1}-A_d| = \frac{\frac{\dd^2}{\dd A^2}\rho_L(\xi)}{\frac{\dd}{\dd A}\rho_L(A_k)}|A_k-A_d|^2$.
By taking an initial amplitude $A_1$ close to $A_d$ such that $\delta |E_1| < 1$ where $\delta = \kappa/\lambda$ with $\kappa=\sup\limits_{\xi\in \mathcal I_E} \left| \frac{\dd^2}{\dd A^2}\rho_L(\xi)\right|$, $\lambda=\sup\limits_{\xi\in \mathcal I_E }\left|\frac{\dd}{\dd A}\rho(\xi)\right|$, 
with $\mathcal I_E := [A_d-|E_1|, A_d + |E_1|]\subset \mathcal I$, 
we can guarantee the quadratic convergence of the iterative remnant update law \eqref{eq:Newton_method} so that the inequality \eqref{eq:quadratic_conv} holds.
\end{proof}

The update law as in \eqref{eq:Newton_method} is akin to the Newton's method. As it may be difficult to obtain numerically the gradient of the remnant curve $\rho_{L}$ at each iteration, one can use an approximation of $\frac{\dd}{\dd A}\rho_{L}$ as follows, 
$\frac{\dd}{\dd A}\rho_{L}(A_k) \approx \frac{\rho_{L}(A_k)-\rho_{L}(A_{k-1})}{A_k-A_{k-1}}$,
which is also known in the literature as the secant method. 
In this case, the secant-method iterative remnant update law is given by
\begin{equation}\label{eq:secant_method}
A_{k+1} = A_k - \frac{A_k-A_{k-1}}{\rho_{L}(A_k)-\rho_{L}(A_{k-1})}e_k.
\end{equation}

For both methods, when the underlying weight $w$ is approximately constant, i.e. $w(\alpha,\beta)\approx C$ with $C>0$, it follows from the computation in the proof of Theorem \ref{thm:Newton_remnant} that $\frac{\dd}{\dd A}\rho_L(A)\approx C A_{\text{max}}$ and $\frac{\dd^2}{\dd A^2}\rho_L(A) \approx 0$ since $\frac{\dd}{\dd A}w(A,\beta)\approx 0$. In this case, we can admit initial amplitude $A_1$ that is far from $A_d$. This particular case is the one considered in the simulation results below.  

 As discussed earlier, we can also consider the complementary reset sub-signals in \eqref{eq:general_iterative_remnant_input} using its signed reverse form. Similar results as in Theorem \ref{thm:Newton_remnant} can be obtained using the same Newton's iterative remnant control law \eqref{eq:Newton_method} and its secant-method one in \eqref{eq:secant_method}.


\section{Numerical simulation}\label{sec:simulation}
\begin{figure}[h]
    \centering
    \subfloat[Histogram]{\label{fig:Histogram}
    \includegraphics[width=0.3\textwidth]{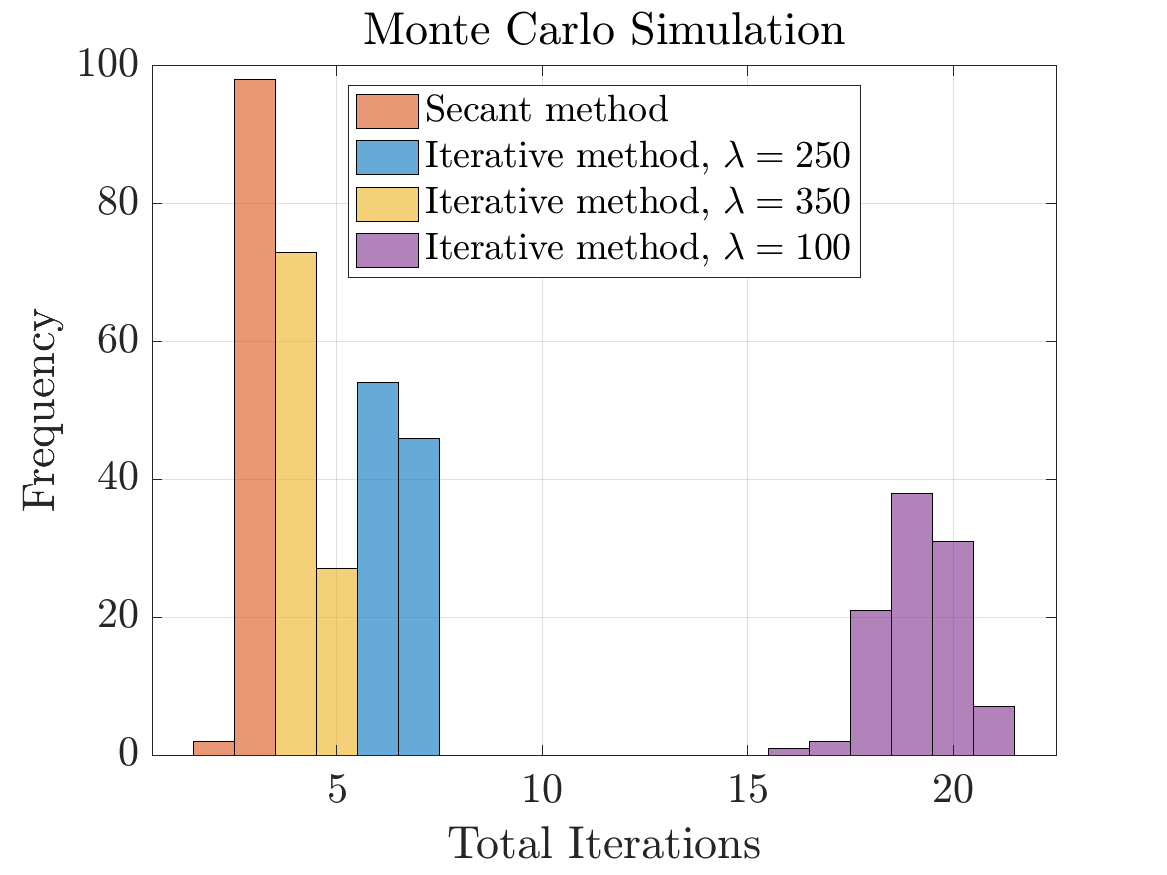}
    }\\
    \subfloat[Transient response]{\label{fig:transient}
    \includegraphics[width=0.3\textwidth]{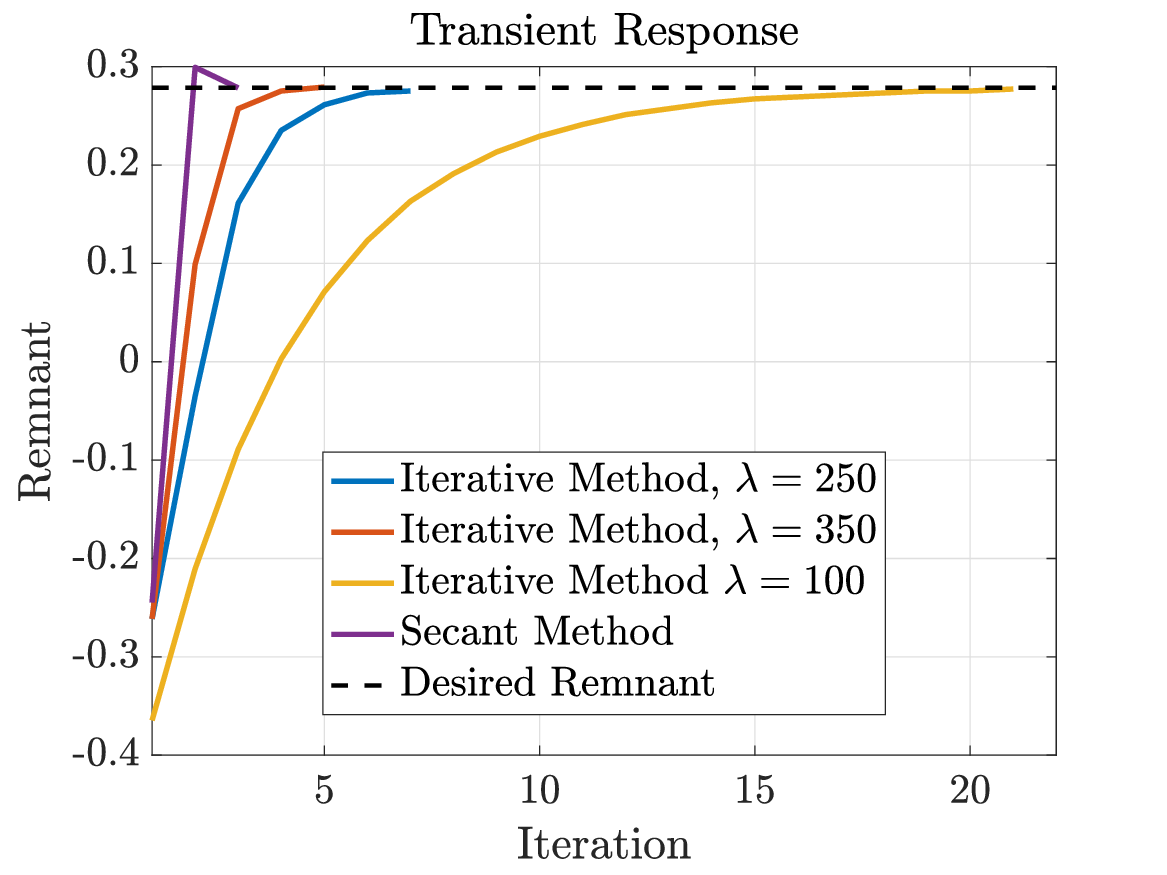}
    }
    \caption{Numerical simulation results, where the proposed secant-based iterative remnant control method is compared with the iterative remnant control method of \cite{vasquez2020recursive} with different values for $\lambda$.}
    \label{fig:Numericsimulation}
\end{figure}

In this section, we perform  numerical simulations of the Newton-based iterative remnant control to track a desired remnant output. We compare the transient response of 
Secant-based method \eqref{eq:secant_method} to 
the iterative control algorithm of \cite{vasquez2020recursive}. 
For the simulation setup, we consider the Preisach operator as in \eqref{eq:Preisach} with randomly generated initial interface $L_0$ and with a uniform weight function $w$. For numerical purpose, we discretize the Preisach plane into 
$N=500500$ regularly spaced relays in the $2$D plane of $[-400,400]\times [-400,400]$ with uniform constant weight of 
$1/N$. 
For the Monte Carlo simulation, we use $100$ samples for each method, where a normally distributed desired remnant output with mean value of $0.1$ and standard deviation of 0.0878 is considered to evaluate the efficacy of the proposed methods in tracking different desired remnant output values.  
We set $A_{\text{max}}=400$ which follows the considered Preisach plane. The initial amplitudes for the secant method are given by 
$A_0=50$ and $A_1=100$, and the initial amplitude for the iterative requires is set at 
$A_0=50$. In addition, three different gains are used for the original iterative remnant control method, to evaluate the influence of $\lambda$ on the convergence rate. 
The resulting Monte Carlo simulation is shown in Figure \ref{fig:Numericsimulation} which shows a histogram of required iterations to converge to the randomly generated desired remnant output and the transient response of sample 1. 
This Monte Carlo simulation result shows 
that the secant-based iterative remnant control method converges faster in all cases to the desired remnant output than the previously proposed iterative remnant control method in \cite{vasquez2020recursive}.
\section{Conclusions}\label{sec:conclusion}
In this paper, we studied the mathematical properties of the remnant curve and showed how these properties can be used to propose a new iterative remnant control law based on Newton and secant methods. 
Further works are underway on the generalization of the remnant control and its analysis to the hysteresis with butterfly loops as studied in \cite{jayawardhana2018modeling}, where monotonicity of the remnant curve is no longer guaranteed. 





\bibliographystyle{IEEEtran}

\bibliography{ref}

\end{document}